\documentstyle[aasms4,epsf]{article}

\begin{document}
\def\rpcomm#1{{\bf COMMENT by RP:  #1} \message{#1}}
\def\uptilde{\mathaccent"164}
\def\etal{{\it et al.~}} 
\def\ls{\vskip 12.045pt}   %One line space.%
\def\ni{\noindent}        %No Indent%
\def\kms{km\thinspace s$^{-1}$ }     %kms -1%
\def\amm{\AA\thinspace mm$^{-1}$ }     %A mm-1%
\def\deg{\ifmmode^\circ _\cdot\else$^\circ _ \cdot$\fi }    %Degree sign%
\def\degg{\ifmmode^\circ \else$^\circ $\fi } 
\def\solar{\ifmmode_{\mathord\odot}\else$_{\mathord\odot}$\fi} %Masses of the sun%
\def\arcs{\ifmmode {'' }\else $'' $\fi}     %Arc seconds%
\def\arcm{\ifmmode {' }\else $' $\fi}     %Arc minutes%
\def\buildrel#1\over#2{\mathrel{\mathop{\null#2}\limits^{#1}}}
\def\mper{\ifmmode \buildrel m\over . \else $\buildrel m\over .$\fi}
%Superscript 'm' over period.%
\def\hper{\ifmmode \rlap.^{h}\else $\rlap{.}^h$\fi}
%Superscript 'h' over period%
\def\sper{\ifmmode \rlap.^{s}\else $\rlap{.}^s$\fi}
%Superscript 's' over period.%
\def\arcsper{\ifmmode \rlap.{' }\else $\rlap{.}' $\fi}
% ' over period.%
\def\arcmper{\ifmmode \rlap.{'' }\else $\rlap{.}'' $\fi}
% " over period.%
\def\gapprox{$_ >\atop{^\sim}$}     %Greater than over approximately (wiggle).%
\def\ltapprox{$_ <\atop{^\sim}$}     %Less than over approximately.%
\def\tworule{\noalign{\medskip\hrule\smallskip\hrule\medskip}}
%Double rule.%
\def\onerule{\noalign{\medskip\hrule\medskip}}
%Single Rule%
\def\et{{\it et~al.~}}
%-----------less/greater than approx eq to--------------
\newcommand{\lta}{{\small\raisebox{-0.6ex}{$\,\stackrel
{\raisebox{-.2ex}{$\textstyle <$}}{\sim}\,$}}}
\newcommand{\gta}{{\small\raisebox{-0.6ex}{$\,\stackrel
{\raisebox{-.2ex}{$\textstyle >$}}{\sim}\,$}}}
\newcommand{\apb}{$(A+B)/2$ }
\newcommand{\amb}{$(A-B)/2$ }

%-------------------------------------------------------
%--------------%
%REFERENCES FOR PUB. A.S.P. CONFERENCE SERIES%
%--------------%
\def\apj{ApJ}  %Astrophysical Journal%
\def\apjs{Ap.~J.~Suppl. }  %Astrophysical Journal Supplements%
\def\apjl{Ap.~J.~ } %Astrophysical Journal Letters%
\def\pasp{{Pub.~A.S.P.} }      %Publications of the Astronomical%
                                %Society of the Pacific%
\def\mn{MNRAS}      %Monthly Notices of the Royal%
                                %Astronomical Society%
\def\aa{Astr.~Ap. }     %Astronomy & Astrophysics%
\def\aasup{AAS }     %A & A Supplements%
\def\baas{Bull.~A.A.S. }  % Bulletin of the American A.S.%

%\lefthead{Guti\'errez et al.}
%\righthead{New CMB structures detected with the Tenerife experiments}

\title{The estimation of the noise in cosmic microwave \\background anisotropy experiments}
\author{C. M. Guti\'errez\altaffilmark{1}}

\altaffiltext{1}{Instituto de Astrof\'\i sica de Canarias, V\'\i a L\'actea s/n, 38200 La Laguna, SPAIN; cgc@iac.es}

\begin{abstract}

Even the most sensitive cosmic microwave background anisotropy
experiments have signal to noise ratios $\lta 5$, so that an accurate
determination of the properties of the cosmological signal requires a
careful assessment of the experimental noise. Most of the experiments
combine simultaneous multi-channel observations in which the presence
of correlated noise is likely. This case is common for ground-based
experiments in which an important fraction of the noise could be
atmospheric in origin. Here, the way to compute and
determine the effects produced by this correlated noise is discussed; in particular,
the paper considers the Tenerife experiments (three radiometers at 10,
15 and 33 GHz with two independent receivers each) showing how this
effect has been taking into account properly in the more recent
analysis of these data. It will be demonstrated that for each of the
three radiometers of these experiments, the atmospheric noise is
equivalent to a Gaussian noise common to both channels with a coherence
time smaller than the binning time, the net effect being an enhancement
of the error-bars in the stacked scan as compared with the estimation
for the case of pure uncorrelated noise. As expected from the spectral
index of the atmosphere, the effect is more important at higher
frequencies. The formalism  is generalized and applied to the general
case of simultaneous multi-channel observations.

\end{abstract}


\section{INTRODUCTION}

The sensitivity of the observations of cosmic microwave background
(CMB) anisotropies for existing and planned experiments will allow
differentiation among theoretical models determining the  parameters of
cosmological interest with an uncertainty of only a few per cent.
However, even the most sensitive experiments have poor signal to noise
ratios (Hancock \et 1994; Bennett \et 1996, etc.) and are sensitive
only to the most intense structures in the CMB signals. The statistical
properties of the CMB signal are obtained by using sophisticated
techniques (likelihood analysis, etc.), which compute the correlations
in the data and compare them with the expectations assuming a given
model. This requires an accurate estimation of all the sources of
correlated and uncorrelated noise (Wilkinson 1995). Lineweaver \et
(1994) and Dodelson, Kosowsky \& Myers (1995) computed the correlated
part of the noise in the COBE DMR (Smoot \et 1992) and Saskatoon data
(Wollack \et 1993), respectively. This paper is mainly concerned with
the Tenerife beam-switching radiometers (Hancock \et 1994, 1996;
Guti\'errez \et 1995; Davies \et 1996). These experiments cover an
angular range (multipoles $l=10-40$) of great interest for establishing
the spectral index of the primordial fluctuations and the total density
of the Universe and for differentiating between scalar and tensorial
modes. No other experiments in this angular range are planned for the
near future. Improvements on the results of the Tenerife experiments
will require deeper integration time and an extension of the region
observed in the sky, but also better estimations of the contaminating
foregrounds and the assessment of the atmospheric contribution.  This
paper focuses on the way in which this atmospheric noise can be
evaluated. Section 2 is dedicated to addressing this problem for the
Tenerife experiments, whilst Section 3 presents the formalism for the
general case of experiments with $n$ channels at different frequencies.
Finally Section 4 presents the conclusions.

\section{APPLICATION TO THE TENERIFE EXPERIMENTS}

The experiments consist of three radiometers operating at 10, 15 and 33
GHz (two independent receivers each) installed on Tenerife at an
altitude of 2400 m. For more than ten years the instruments have
collected data in seven bands separated 2.5\degg~from Dec.=+30\degg~to
Dec.=+45\degg, scanning a band at constant declination each day.
Detailed descriptions of the instruments, observing technique and data
processing can be found in Davies \et (1996). The analysis presented
here is dedicated to the data at Dec.=+40\degg~presented in Hancock \et
(1994, 1996), however the conclusions are similar to those obtained
with the more recent data at Dec.=+35\degg~(Guti\'errez \et 1997).
 The noise in a beam-sized element ($\sim
$5\degg) in each of the individual observations are $\sim 0.6$, $\sim
0.4$, and $\sim 0.3$ mK at 10, 15 and 33 GHz respectively, whilst the expected
amplitudes of the CMB signal for the instrumental configuration is
$\sim 30-40$ $\mu$K; this makes it necessary to repeat the observations $\sim 100$ times in order to achieve enough sensitivity in the CMB signals.
In the data presented by Hancock \et (1994) each position in RA
correspond to a mean number of 120, 130 and 85 observations at 10, 15
and 33 GHz, respectively. The estimation of the error-bars in the
stacking presented by Hancock \et (1994) was done assuming that the noise
in each individual observation were purely uncorrelated. This is
correct for the instrumental noise, which has a thermal origin; however,
most of the observations in each instrument correspond to measurements
with both channels simultaneously (there is a small time delay of 16
ms between them), and therefore any possible atmospheric contribution
will introduce correlation between both channels.

This amplitude and temporal scale of coherence of this effect have been
evaluated studying the correlation and cross-correlation functions
between channels.  Figure 1 represents the correlation of the observations
as a function of time (which is equivalent to separation in RA). The
plots are an extension of Figure 7 in Davies \et (1996). The top and middle panels
indicate that the sources of noise are uncorrelated on scales larger
than 4 min (the binning time). The cross-correlations plotted at the bottom show that a part of the
noise is common to both channels; it is likely that the atmosphere is
the source of this correlated noise. The distribution of the data in
each individual scan has been studied using the Kolmogorov-Smirnov test
which shows the compatibility with a Gaussian distribution. In summary, 
these analyses indicate that the effect of the atmosphere on the
individual observations can be modelled as a Gaussian noise common to
simultaneous measurements of both channels with a time coherence scale
smaller than 4 min, and therefore independent between adjacent
positions in RA. As expected from the spectral index of atmospheric
emission, the largest effect is at 33 GHz. The amplitude of this
atmospheric noise can be inferred from the cross-correlation between
both channels. I made Monte Carlo simulations of the data and compute
the auto-correlation and cross-correlation in the same way as the
actual data. The results of these simulations (dashed lines in Figure 1) show good agreement with the actual
correlations functions, demonstrating the validity of the model.

Denoting the atmospheric signal by $a$ and the signal due to the instrumental
noise by $r_1$ and $r_2$
($<r_1^2>\approx <r_2^2>$), in channels 1 and 2 respectively, combining the results of both
channels the variance of the stacked scan is
\begin{equation}
<(\frac{1}{2}[(a+r_1)+(a+r_2)]^2>=\frac{1}{4}\{4<a^2>+<r_1^2>+<r_2^2>\}.
\end{equation} However, ignoring the correlation due to the
atmosphere we would obtain (incorrectly) $\frac{1}{4}\{2<a^2>+<r_1^2>+<r_2^2>\}$
and therefore the noise would be underestimated  by a factor
$g_{+}=\sqrt{1+\epsilon}$, where
\begin{equation}\epsilon =\frac{1}{1+(<r_1^2>+<r_2^2>)/2<a^2>}.
\end{equation}
This is an extension of expression (1) in Davies \et (1996) for the
case in which the instrumental noise in both channels is not exactly
the same. In CMB experiments which consist of two channels ($A$ and $B$) usually the
signal is computed comparing the sum $(A+B)/2$ (which contains the
signal plus the noise) and the difference $(A-B)/2$ (which only
contains the noise see Hancock \et 1994). In this case, is easy to
show that, ignoring the correlated term, there is an overestimation of
the noise in the $(A-B)/2$ scan by a  factor
$g_{-}=\sqrt{1-\epsilon}$.  This factor is a consequence of the
subtraction of part of the noise (the atmospheric part) when the
difference $(A-B)/2$~scan is computed.  In the limit in which the
atmospheric noise is negligible as compared with the instrumental
noise, we obtain $g_{+}=g_{-}=1$, as expected; in the case in which the
atmospheric noise is much more larger than the instrumental noise we
obtain $g_{+}\rightarrow \sqrt{2}$ and $g_{-}\rightarrow1/\sqrt{2}$ and
then simultaneous observations with two channels are fully redundant.

For the Tenerife CMB data at Dec=+40\degg, it is possible to distinguish
between the instrumental and the atmospheric noise by analyzing the
distribution of the cross-correlation between both channels, as
explained above. The deduced $g_{+}$ and $g _{-}$ factors have the
values quoted in Table~1. As expected, the table indicates that the data
which are more affected are the 33 GHz. The mean correlation between
channels is larger at 15 GHz than at 10 GHz as can be seen in Figure 1,
however in Table ~1 the $g_{+,-}$ factors are similar for both
instruments; the reason for this is that one of the channels at 15 GHz
was not operating due to a malfunction most of the time. The data at 15
and 33 GHz were recorded in different campaigns and therefore there
are not an extra factor due to the correlated part between data at
these two frequencies.

How this re-estimation of the error-bars affects the analyses of the
estimation of the CMB signal? Table~1 of Hancock \et (1994) computed the
rms of the astronomical signal ($\sigma _{\rm RMS}$) present at each
frequency from the signals in the $(A+B)/2$ ($\sigma_{(A+B)/2}$) and
$(A+B)/2$ ($\sigma_{(A-B)/2}$) scans, $\sigma
_{\rm RMS}^2=\sigma_{(A+B)/2}^2-\sigma_{(A-B)/2}^2$. As has been shown
above, these values of $\sigma_{(A+B)/2}$ and $\sigma_{(A-B)/2}$ are affected by the atmospheric signal
common to both channels. Following the formalism explained above, I have computed the atmospheric $\sigma _a$ and astronomical signals $\sigma _s$ presented
in each of the final stacked scans and corrected the values quoted in
Table~1 of Hancock \et (1994). The new and old ($\sigma _{old}$) values are presented in columns 4, 5 and 6 of Table 1.

The correlated noise also affects to the likelihood analysis. The
re-estimation of the error-bars presented above is equivalent to
increasing the diagonal term in the covariance matrix. Figure 2
presents the likelihood curves for the 15, 33 and combined 15+33 data
with the old ($a$, $b$ and $c$) and the new ($d$, $e$, $f$)  estimation
of the noise in the case of a Harrison-Zel'dovich spectrum for the
primordial fluctuations. All the cases show evidence of a clear
well-defined maximum and similar shapes after the re-estimation of the
error-bars.  However, the value of the peak is smaller with the new
estimation. For instance, in the 15+33 data the likelihood peak
normalized with respect to the value for zero signal changes from $\sim
5\times 10^6$ to $5\times 10^4$. This indicates a lower significance of
the detection. Using a Bayessian analysis for the 15+33 data and
assuming a Harrison-Zel'dovich for the spectrum of the primordial
fluctuations, a normalization of $Q_{\rm RMS-PS}=26\pm 6$
$\mu$K was obtained; with the new estimation of the error-bars in the stacked data
it is obtained $Q_{\rm RMS-PS}=22^{+10}_{-6}$ $\mu$K. This detection is
still very significant and it would be necessary to increase the
error-bars artificially by a factor as large as $\sim 2.6$ to make the
likelihood curve compatible with pure noise. Figure 2 also presents
(panels $g$, $h$ and $i$) the likelihood curves of the $(A-B)/2$ with
the new error-bars. In the three cases the scans are compatible with
noise, emphasyzing that the detected signals are common to both
channels and frequencies.

\section{GENERAL CASE}

The formalism presented in last section can be easily generalized for
any other experiment with $n$ simultaneous channels operating at
different frequencies.  Specifically for experiments working at mm
wavelengths, it is expected that atmospheric noise would be the main
source of noise (Piccirillo \et 1997). Following the results of the
previous section, it will be considered an experiment in which the
contribution due to the atmosphere is equivalent to Gaussian noise
uncorrelated point to point with a certain dependence in frequency. The
result in each channel consists of the signal on the sky plus two
components, the first due to thermal uncorrelated noise and the second
which is correlated in some degree between the different channels. I
assume that the noise in the i-$th$ channel is $k_ia+r_i$, where $k_ia$
is the contribution of the atmosphere which has a spectral dependence
$k_i$, and $r_i$ is the part due to the thermal noise. When the results of the $n$ channels are combined, the variance of the stacked scan is
$$<(\frac{1}{n}\sum _i(k_ia+r_i)^2>=
\frac{1}{n^2}[\sum_i<(k_ia+r_i)^2>+
2\sum_{i,j>i}<(k_ia+r_i)(k_ja+r_j)>] $$
\begin{equation}
=\frac{1}{n^2}\{<a^2>\sum _i k_i^2+\sum _i<r_i^2>+2\sum_{i,j>i}k_i k_j<a^2>\}.
\end{equation}
However, if we ignore the correlation due to the atmosphere we would incorrectly obtain
\begin{equation}
\frac{1}{n^2}\{<a^2>\sum_ik_i^2+\sum _i<r_i^2>\}
\end{equation}
and therefore we would underestimate the noise by a factor
\begin{equation}
g _{+}=\sqrt{1+\frac{2\sum_{i,j>i}k_i k_j}{\sum_i k_i ^2+\frac{\sum _i<r_i^2>}{<a^2>}}}.
\end{equation}
Especially if $a=0$, $g=1$, and when $r=0$, $g_{+}
=\sqrt{1+\frac{2\sum _{i,j>i}k_i k_j}{\sum _i k_i^2}}$. In the case of two channels operating at the same frequency, the expression of Section 2 is recovered. The factor $g_{-}$, when we compute the noise in the difference between channels $i$ and $j$, is
\begin{equation}
g_{-}=\sqrt{1-\frac{2k_ik_j<a^2>}{<r_i^2>+<r_j^2>+(k_i^2+k_j^2)<a^2>}}.
\end{equation}

\section{CONCLUSIONS}

In this paper it has been shown how the presence of correlated noise affects
the estimation of the error-bars when data from several channels are
combined. This is particularly interesting in cases of poor signal-to
noise-ratios like the CMB observations. It has been shown that for the
Tenerife experiments the atmosphere can be modeled like Gaussian
noise uncorrelated between adjacent measurements. The net effect, in
this case, is an enhancement of the error-bars in the final stacked
scans. The maximum effect is at 33 GHz where more than half of the
noise is atmospheric in origin. With the new estimation of the noise,
the likelihood analysis of the Tenerife data at Dec.=+40\degg~shows the
presence of clear CMB signals corresponding to an expected quadrupole
$Q_{\rm RMS-PS}=22^{+10}_{-6}$ $\mu$K, a value slightly smaller in amplitude
and in statistical significance than the old estimation presented in
Hancock \et (1994). Obviously, the effect is highly dependent on the
observing technique; for instance the preliminary analysis of the data of the Tenerife
interferometer at 33 GHz does not show evidence of correlated noise
between the sine and cosine channels, even in cases of relatively bad
weather. I have assumed here that the
origin of the correlated noise between channels is the atmosphere, but the
formalism can be applied independently of the physical
origin of the noise.

\subsection*{ACKNOWLEDGMENTS}

\noindent I am grateful to the members of the Tenerife experiments for access to the data analyzed here, and for useful discussions and comments. 

{}

\clearpage
\subsection*{FIGURE CAPTIONS}
 
\figcaption[fig1.eps]{The correlation of the signals in the Tenerife CMB data  as a function
of the shift in RA. Panels at the top, middle and bottom are for the
auto-correlation of channels 1 and 2, and the cross-correlation between
them, respectively. The points correspond to the mean of the
correlation function and the solid lines enclose the 68 \% c.l. along
the different observations. The dashed lines correspond to Monte Carlo simulations of the data.} 

\figcaption[fig2.eps]{Likelihood functions of the results at 15 GHz (left), 33 GHz (middle) and the weighted addition of both 15+33 (right). Panels at the top show the likelihood function of the signal in the $(A+B)/2$ without considering the presence of correlated noise between channels. Plotted in the second line are the new likelihood curves after the re-estimation of the error-bars. The likelihood curves for the $(A-B)/2$ scans are at the bottom.}

\newpage

\begin{table}[thb]
\begin{center}
\begin{tabular}{lccccc}
Frequency & $g _{+}$ & $g _{-}$ & $\sigma _a$ ($\mu$K) & $\sigma _s$ ($\mu$K) & $\sigma _{old}$ ($\mu$K) \\
\hline
\hline
\\
10 GHz& 1.03& 0.96 & \dots & 36 & 36 \\
\\
15 GHz & 1.04& 0.96 & 23 & 34 & 41 \\
\\
33 GHz& 1.16& 0.86 & 32 & 37 & 49 \\
\\
15+33 & 1.12 & 0.89 & 22 & 36 & 42\\
\end{tabular}
\end{center}
\caption{Corrections for the Tenerife results.}
\end{table}

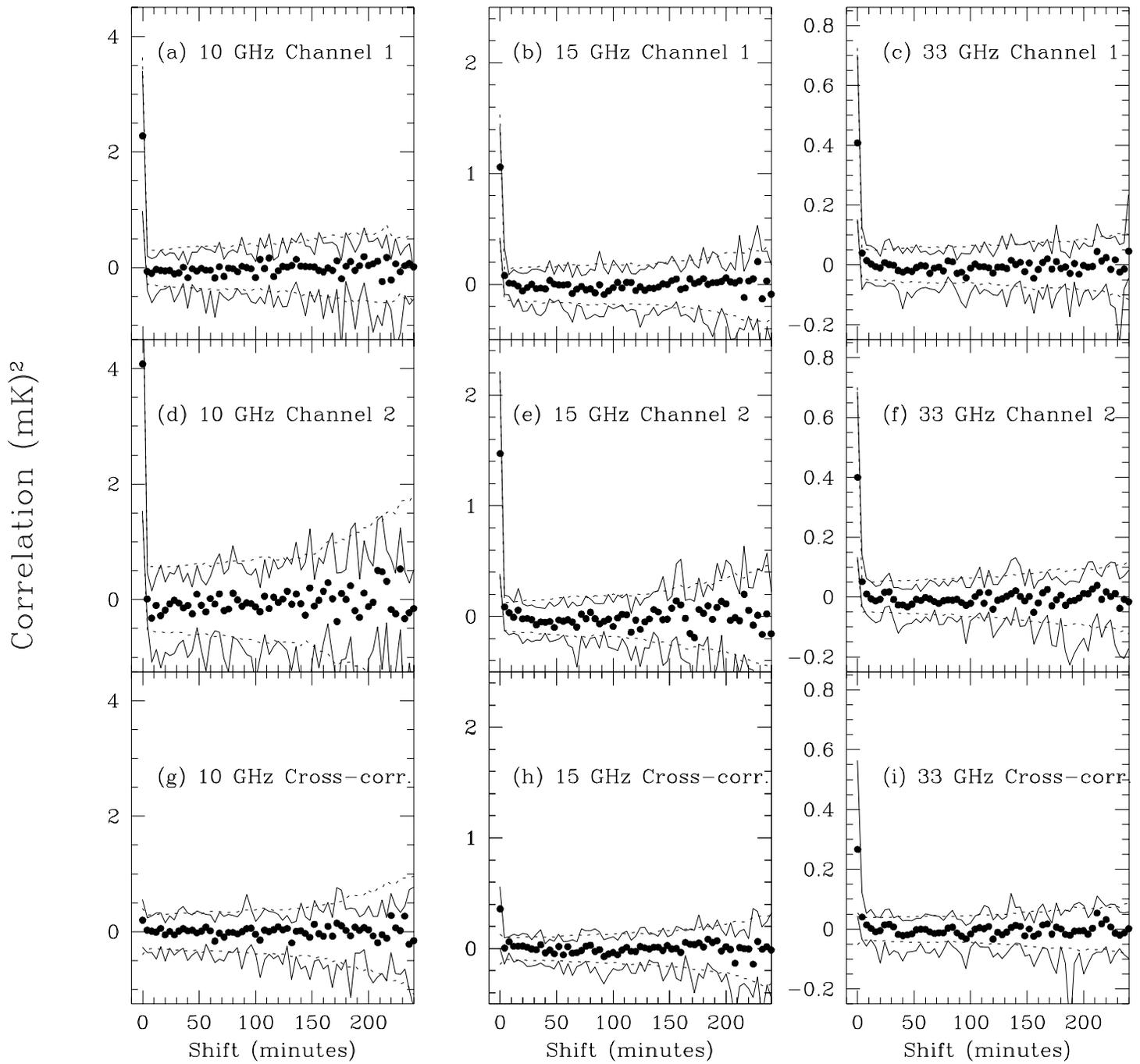
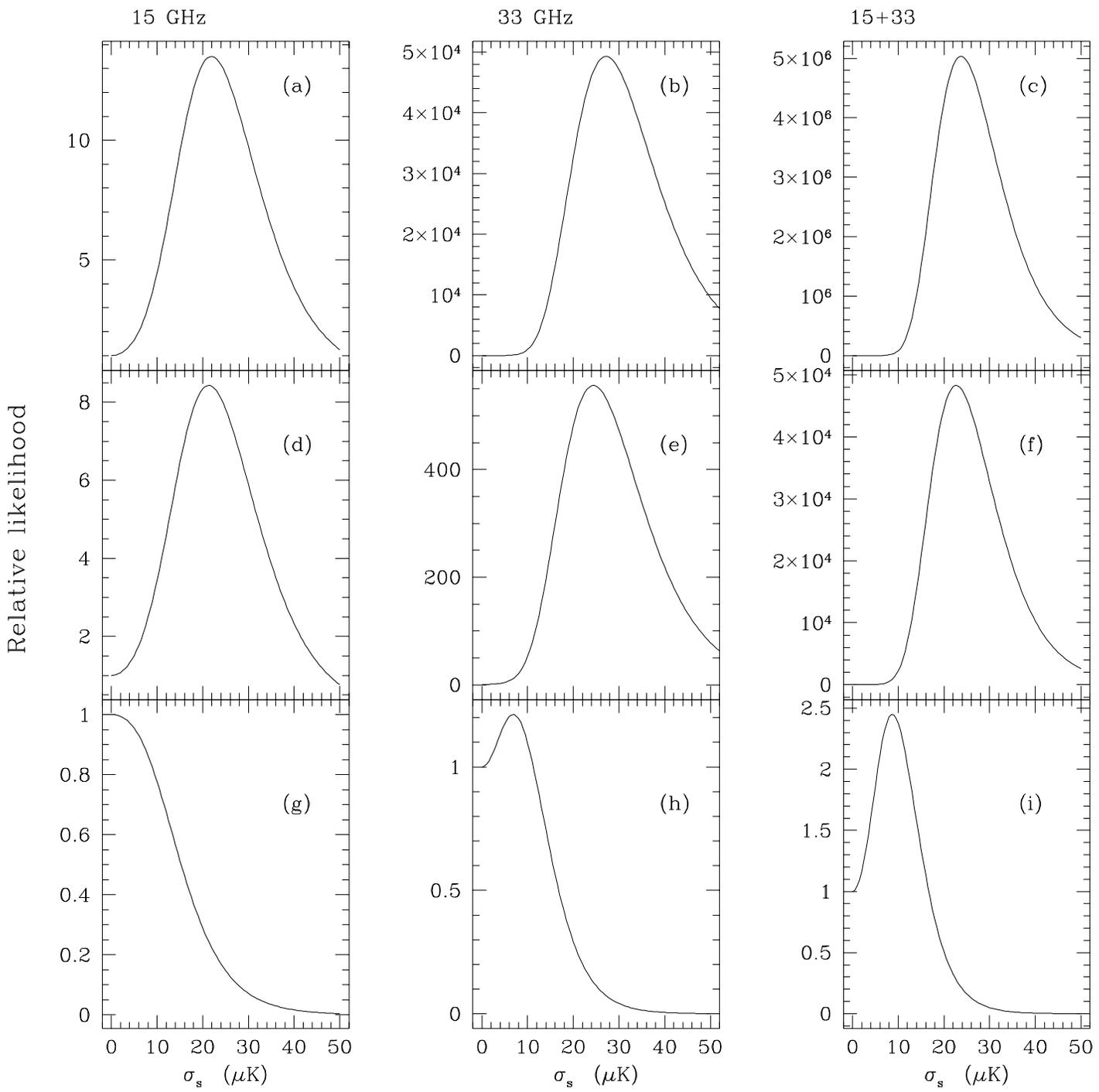
\end{document}